\documentclass{ws-ijmpb}

\begin{document}

\markboth{Y. Barlas, W. C. Lee, K. Nomura, A. H. MacDonald}
{Renormalized Landau Levels and Particle-hole symmetry in Graphene}

%
\catchline{}{}{}{}{}
%

\title{RENORMALIZED LANDAU LEVELS AND PARTICLE-HOLE SYMMETRY IN GRAPHENE}

\author{YAFIS BARLAS}

\address{Physics Department, University of Texas at Austin, 1 University Station\\
Austin, Texas, 78712, USA\\
National High Magnetic Field Laboratory, Florida State University, 1800 E Paul Dirac 
Dr\\
Tallahassee, Florida, 32310, USA 
barlas@magnet.fsu.edu}

\author{WEI-CHENG LEE}

\address{Physics Department, University of California at San Diego, 9500 Gilman Drive\\
La Jolla, California,  92093, USA
leewc@physics.ucsd.edu}

\author{KENTATO NOMURA}
\address{Department of Physics, Tohoku University \\
Sendai, 980-8578, Japan
nomura@cmpt.phys.tohoku.ac.jp}

\author{ALLAN H MACDONALD}
\address{Physics Department, University of Texas at Austin, 1 University Station\\
Austin, Texas, 78712, USA\\
macd@physics.utexas.edu}

\maketitle

\begin{history}
\received{Day Month Year}
\revised{Day Month Year}
\end{history}

\begin{abstract}
In this preceedings paper we report on a calculation of graphene's Landau levels in a 
magnetic field. Our calculations are based on a self-consistent Hartree-Fock 
approximation for graphene's massless-Dirac continuum model.  We find that 
because of graphene's chiral band structure interactions not only shift 
Landau-level energies, as in a non-relativistic electron gas, 
but also alter Landau level wavefunctions.  We comment on the subtle continuum 
model regularization procedure necessary to correctly 
maintain the lattice-model's particle hole symmetry properties. 
\end{abstract}


\section{Introduction}

Graphene, a one-atom-thick two-dimensional crystal of carbon atoms arranged 
in a honeycomb lattice, is a gapless semiconductor with an unusual massless 
Dirac-fermion band structure that has long attracted theoretical 
interest~\cite{semenoff,haldane}. The  
low-energy properties of graphene are characterized by 
quasiparticle dispersion~\cite{earlyguy} linear in momentum and by 
vanishing density-of-states at the neutral system Fermi energy.
The band eigenstates can be considiered as sublattice-pseudospin 
spinors and have a chiral property which 
qualitatively alters the way in which electron-electron interactions influence 
electronic  properties. In particular electron-electron interactions lead to a
logarithimic enhancement of the Fermi velocity in doped and undoped graphene
related to a lack of screening at the Dirac 
point~\cite{gin,polinissc,vafek}.\\ 
In the presence of a magnetic field, graphene's electronic 
structure  also changes in a nontrivial way when compared to the non-relativistic 
two dimensional electron 
gas (2DEG) case, leading to the so-called half-quantized Hall effect~\cite{qhtheory,qhexpt}
in which the plateau values of the  
Hall conductivity are given by $\sigma_{xy} = 4(n+1/2)(e^{2}/h)$.
Plateau conductivity values are separated 
by $4e^{2}/h$ because of the fourfold degeneracy due to valley and spin. In this paper we analyze the effect 
of electron-electron interactions on graphene's LL spectrum.  We show that because of the 
chiral nature of graphene's band structure, interactions not-only shift Landau level
energies but also alter Landau-level wavefunctions. 

\section{Self Consistent Hartree-Fock Approximation}

In this section we use the self-consistent Hartree-Fock approximation (SCHFA) to study the 
effect of electron-electron interactions on the LL spectium of graphene within the 
massless Dirac-fermion(MDF) model.  
The low-energy properties of graphene can be adequately described by a MDF model:
\begin{equation}
\label{MDF}
\mathcal{H}_{\vec{p}} = v \vec{\sigma} \cdot p
\end{equation}
where $\sigma^{i}$'s are Pauli matrices acting on graphene's psuedospin degrees of 
freedom, $\vec{p}$ is a two dimensional vector 
measured from the Dirac points.  (As we will discuss later this model requires 
especially subtle ultraviolet regularizations in order to yield physically correct 
predictions.) 
In the presence of a unifrom magnetic field $\vec{B} = 
B \hat{z}$ applied in a 
direction perpendicular to the plane of the graphene sheet~\ref{MDF} is modified by 
 $\vec{p} \to \vec{\pi} = \vec{p} -(e/c)\vec{A}$ where $\vec{A}$ is the vector 
potential with $\vec{B} =\vec{\nabla} \times \vec{A}$. Defining the usual raising and 
lowering LL operators $\vec{a^{\dagger}}$ and $a$, with $a^{\dagger} = 
(l_{B}/\sqrt{2}\hbar) \pi$, where $l_{B} = (\hbar c/eB)^{1/2} = 
25.66/\sqrt{(B}[\rm{T}])\rm{nm}$ is the magnetic length, we can identify a 
zero-energy eigenstate given by $a \phi_{0} = 0$ and finite-energy chiral 
eigenatates $n$ and 
$\bar{n} = -n$ with eigenenergies $\varepsilon_{n} = sgn(n) \sqrt{\frac{2 \hbar v^{2}e 
B |n|}{c}}$.  In the Landau gauge $ \vec{A} = (0,Bx,0 
)$, the corresponding eigenvectors are
\begin{equation}
|\psi_{n \neq 0 , X} \rangle  = \frac{1}{\sqrt{2}}\left(
\begin{array}{c} -i sgn(n) \phi_{|n|-1,X}\\
\phi_{|n|,X} \end{array} \right),
\qquad
|\psi_{n = 0 , X} \rangle  = \left(
\begin{array}{c} 0 \\
\phi_{0,X}\end{array} \right),
\end{equation}
where $\phi_{n,X}$ is a Landau-gauge eigenstate of a non-relatvistic electron gas and 
$X$ denotes the guiding center degree of freedom within a LL.  Projecting the 
interacting Hamiltonian onto the LL basis gives: 
\begin{equation}
\mathcal{H}_{e-e} = \frac{1}{2} \sum_{\vec{q}} v_{q} : \hat{\rho}(-\vec{q}) 
\hat{\rho}(\vec{q}):,
\end{equation}
where $v_{q} = 2 \pi e^{2}/ \epsilon |q|$ is the two-dimensional Fourier transform of 
the Coulomb interaction and
\begin{equation}
\hat{\rho}(-\vec{q}) = \sum_{n X n' X'} c^{\dagger}_{n X \tau s} c_{n' X'
\tau' s'} \delta (q_{y} + X - X') e^{i q_{x}(X'-\frac{q_{y}}{2})}.
F^{R}_{n,n'}(\vec{q}),
\label{density}
\end{equation}
In Eq.(~\ref{density}) 
$c^{\dagger}_{n X \tau s} /c_{n X \tau s}$ are creation/annihilation 
operator for particles in LL $n$ at guiding center $X$ for valley $\tau $ and spin
$s$. For notational simplicity we have assumed $l_{B} =1$; we will 
however restore these length units in the final results. $F^{R}_{n,n'}(\vec{q})$ 
is referred to as 
graphene's relativistic form factor and captures the orbital and 
sublattice pseudospin character of the LL 
orbitals~\cite{kentaroQHF} 
($n \neq 0 $ and $n' \neq 0$): 
\begin{equation}
F^{R}_{n,n'}(\vec{q}) = \frac{1}{2} \big[ F_{|n|,|n'|}(\vec{q}) + sgn(nn')
F_{|n|-1,|n'|-1}(\vec{q}) \big],
\end{equation}
$ F_{|n|,|n'|}(\vec{q})$ is the well known form factor for an ordinary 
2DEG~\cite{allanearlypapers} in the presence of a perpendicular magnetic field:
\begin{equation}
F_{n,n'}(\vec{q}) = \bigg\{ \begin{array}{cc}
\sqrt{\frac{n'!}{n!}} \big[i\frac{q_{x} -i q_{y}}{\sqrt{2}} \big]^{n-n'}
L_{n'}^{n-n'}(\frac{q^{2}}{2}) e^{-q^{2}/4} & n \geq  n'\\
\sqrt{\frac{n!}{n'!}} \big[i\frac{q_{x} + i q_{y}}{\sqrt{2}} \big]^{n'-n}
L_{n}^{n'-n}(\frac{q^{2}}{2}) e^{-q^{2}/4} & n' > n,
\end{array}
\end{equation}
and the form factor for the lowest LL is just $F_{00}^{R}(q) = F(q)= e^{-q^{2}/4}$.  Finally 
\begin{equation}
F^{R}_{0,n}(\vec{q}) = \frac{1}{\sqrt{2}}F_{0,n}(\vec{q}) = \sqrt{\frac{1}{2n!}}
\big[i\frac{q_{x} -i
q_{y}}{\sqrt{2}} \big]^{n} L^{n}(\frac{q^{2}}{2}) e^{-q^{2}/4}
\end{equation}
Here $ L_{n'}^{n-n'} $ are the associated Laguerre polynomials. 

In the Hartree-Fock approximation the effective single-particle Hamiltonian depends on the 
density matrix.  In the case of Landau-level systems, the density-matrix is usefully 
parameterized as follows:
\begin{equation}
\Delta^{n,n'}_{\tau, \tau',s,s'} (\vec{q})= \frac{1}{N_{\phi}} \sum_{X,X'} \langle
c^{\dagger}_{n',X', \tau', s'}c_{n,X, \tau, s} \rangle \delta(q_{y}+X'-X).
\exp^{- iq_{x} X}
\end{equation}
The Hartree-Fock theory Hamiltonian is expressed in terms of $\Delta$ as follows:
\footnote{The details of this calculation 
are similar to the one in Ref. 10.} 
\begin{eqnarray}
\label{finalHFham}
\langle n,X,\tau,s | \mathcal{H}_{e-e} |n',X',\tau',s' \rangle &=& 
\sum_{n_{1},n_{2}} \sum_{q}
\bigg[ \frac{1}{2 \pi l_{B}^{2}} V_{n,n',n_{2},n_{1}} (\vec{q})
\Delta^{n_{2},n_{1}}_{\tau", \tau",s",s"}(\vec{q}) \delta^{ss'}_{\tau \tau'} \\
\nonumber
&-& \frac{1}{L^{2}} \sum_{p} V_{n_{2},n',n,n_{1}} (\vec{p}) 
\Delta^{n_{2},n_{1}}_{\tau, \tau',s,s'}(\vec{q}) \exp^{i (\vec{p} \times 
\vec{q})l_{B}^{2} \cdot \hat{z}} \bigg] 
\exp^{i q_{x} X'}
\delta(q_{y}l_{B}^{2} + X - X')
\end{eqnarray}
here we restore $l_{B} $ and define
\begin{equation}
V_{n_{1},n_{2},n_{3},n_{4}}(\vec{q}) = v_{q}
F^{R}_{n_{1},n_{4}}(\vec{q})F^{R}_{n_{2},n_{3}}(-\vec{q})
\end{equation}
For the purposes of this paper we assume that the translational
invariance is not broken.  In this case $\Delta$ is non-zero only for $\vec{q}=0$.
It follows that 
\begin{equation}
\Delta^{n_{1},n_{2}}_{\tau \tau' s s'}(\vec{q}) =\Delta^{n_{1},n_{2}}_{\tau \tau' s
s'}(0) \delta_{\vec{q}=0} \delta(|n_{1}| - |n_{2}|).
\end{equation}
The $\delta$ function which sets mixing between states with 
different values of $|n|$ can be seen to follow from spatial 
isotropy in the continuum model.  Because both positive and negative 
values of $n$ occur in graphene this restriction does not 
forbid mixing of states with positive and negative $n$ by interactions.
The fact that Landau-level wavefunctions are altered by interactions in 
graphene is the main difference between relativistic and non-relativistic 
cases.

Assuming no broken translational symmetry, the
first term in (\ref{finalHFham}) is just the constant Hartree (electrostatic) potential
which can be absorbed in the zero of energy.  Dropping this term yields
\begin{equation}
\langle n,X,\tau,s | \mathcal{H} |n',X',\tau',s' \rangle = - \sum_{|n_{1}| = 
|n_{2}|}
\frac{1}{L^{2}}  \sum_{\vec{q}} v_{q}
F^{R}_{n_{2},n'}(\vec{q})F^{R}_{n,n_{1}}(-\vec{q}) \Delta^{n_{1},n_{2}}_{\tau \tau' s
s'}(0) \delta(X-X')
\end{equation}
The mixing between $n$ and $\bar{n}$ LL is due to the fact that the self-energy 
is not diagonal in the pseudospin as can be seen by examining spatial isotropy 
consequences more closely.  Isotropy is encoded in the form 
factors in the above equation: 
\begin{equation}
F^{R}_{m,n'}(\vec{q})F^{R}_{n,m}(-\vec{q}) =
F^{R}_{m,n'}(\vec{q}) \bar{F}^{R}_{m,n}(\vec{q}) \sim  \exp^{-i \varphi (|n'|-|n|)}
\end{equation}
yielding an angular integral that is proportinal to $\delta(|n'| - |n|) $ not $\delta 
(n'-n)$. The other delta function gives $X=X'$.  Assuming no spin or valley broken 
symmetries we can also assume that 
different valleys can be considered independently.  This yields (suppressing 
the spin and valley indices):
\begin{equation}
\label{finaleeham}
\langle n |\mathcal{H}_{e-e}| n' \rangle = -\sum_{|n_{1}| = |n_{2}|}
\frac{1}{L^{2}}  \sum_{\vec{q}} v_{q}
F^{R}_{n_{2},n'}(\vec{q})F^{R}_{n,n_{1}}(-\vec{q}) \Delta^{n_{1},n_{2}}.
\end{equation}
Notice that this matrix element is non-zero only if $|n|=|n'|$.

To compute the matrix elements we have to employ a LL index cutoff reminisent of the 
high-energy cutoff used in for MDF description of graphene at zero magnetic field.
This ultraviolet cutoff plays a role because of the unbounded negative energy sea of the 
massless Dirac model.  Following the procedure used at zero magnetic field,
we choose a LL cutoff, a maximum value for $|n|$,  
M based on the physically natural cutoff of momentum at an inverse 
lattice constant scale, $k_c \sim 1/a$, and on the semi-classical relationship between 
momentum and Landau-level index.  This yields
\begin{equation}  
\hbar v_{F} k_{c} = \sqrt{2} \frac{\hbar v_{F}}{l_{B}} \sqrt{M}.
\end{equation}
Using $k_{c} \sim 1/a$ where $a= 0.246nm$ is 
graphene honeycomb lattice constant we get a magnetic field dependent cutoff 
\begin{equation}
M \sim \frac{5000}{B[T]}
\end{equation}
where $B$ is the magnitude of the magnetic field. We can 
write the mean field Hartree-Fock hamiltoninan in the $(n,\bar{n})$-sector as a 
two level system:
\begin{equation}
\mathcal{H}_{MF} = \mathcal{E} + \vec{\mathcal{B}} \cdot \vec{\sigma}
\label{hmf}
\end{equation}
where the pseudospin electric $\mathcal{E}$ and magnetic $\mathcal{B}$ field in the 
$(n,\bar{n})$-space $\mathcal{E} = - \alpha_{gr}(\sqrt{\pi}/2)  (V_{n} + 
\bar{V}_{n})/2 $, $\mathcal{B}_{x} =   - \alpha_{gr}(\sqrt{\pi}/2) \tilde{V}_{n}$ 
and $\mathcal{B}_{z} =  \sqrt{n} - \alpha_{gr}(\sqrt{\pi}/2) ((V_{n} - 
\bar{V}_{n})/2)$ ,depend on the interaction matrix elements 

\begin{equation}
V_{n} = \frac{l_{\rm B}}{e^{2}} \sqrt{\frac{2}{\pi}} \langle n| H_{e-e} | n \rangle,
\qquad \bar{V}_{n} = \frac{l_{\rm B}}{e^{2}} \sqrt{\frac{2}{\pi}} \langle \bar{n}|
H_{e-e} | \bar{n} \rangle, \qquad \tilde{V}_{n} = \frac{l_{\rm B}}{e^{2}}
\sqrt{\frac{2}{\pi}} \langle n| H_{e-e} | \bar{n} \rangle,
\label{twobytwo}
\end{equation}
$\alpha_{gr} = e^{2}/(\hbar v_{F})$ is graphene's coupling constant.


\section{Results and Discussions}
\begin{figure}[bt]
\label{figure1}
\centerline{\psfig{file=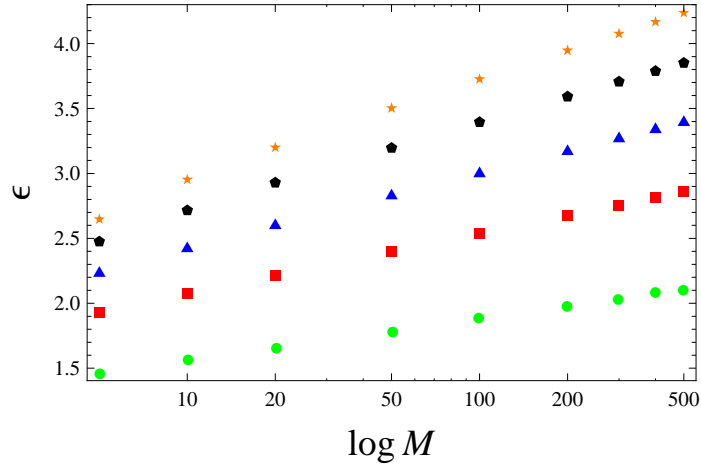,width=3.65in}}
\vspace*{8pt}
\caption{Hartree-Fock eigenvalues for a neutral graphene sheet in the units $\sqrt{2 
\hbar}v/l_{B}$ with $\alpha_{gr}=2.2$. In this calculation the energy 
of the $n=0$ Landau level is chosen as the zero of energy so that all $n \ne 0$ 
eigenstatesoccur in particle-hole symmetric positive and negative energy pairs.}
\end{figure}

We must now address the regularization procedures that are necessary to extract 
useful predictive results from these Hartree-Fock calculations.
The exchange energies $V_{n}$ and $\bar{V}_{n}$ in Eq.(\ref{twobytwo}) both 
diverge with cutoff $M$ like $\sqrt{M}$ while $\mathcal{B}_{z}$ 
diverges like $\ln{M}$ and $\mathcal{B}_{x}$ goes to zero like $1/\sqrt{M}$.
These large exchange energies are indeed physical because graphene's 
$\pi$-electron system has a high density of electrons, close to one per 
honeycomb lattice unit cell.  This large energy is however neither easily 
measureable or of any great interest.  Instead we want to use our 
Hartree-Fock theory to calculate the spacing of energy levels 
near the Dirac point and their dependences on the total Landau
level filling factor.  Progress can be made by simply choosing a 
convenient zero of energy, as we do in the zero-field case.  We propose using 
the energy of the $n=0$ Landau level, $\langle 0 | H_{e-e} | 0\rangle$ 
{\em evaluated for a neutral graphene sheet} as the zero of energy.
In zero field the analogous choice solves all problems, but that is not 
true in the massless Dirac model case as we now discuss.

In order to explain the problem which remains and our resolution of this problem 
we go to a more microscopic level by considering properties not of the
massless Dirac model but of the one-band nearest neighbour tight-binding model
for graphene's $\pi$-orbitals.  This model has the particle-hole symmetry property that when the Hamiltonian acts on a 
wavefunction which is restricted to one sublattice of the bipartite
honeycomb lattice, it produces a wavefunction confined to the other sublattice.
From this property it readily follows that eigenstates of the band 
Hamiltonian occur in positive and negative energy pairs which have 
opposite intersublattice phases and, importantly for the Hatree-Fock 
calculations, that the density-matrix of a neutral graphene sheet is just half
of the trivial density matrix of a state in which all $\pi$-orbital states 
are full,
\begin{equation}
\Delta_{i',i}^{neutral} = \frac{1}{2} \; \delta_{i',i'} .
\end{equation} 
This property is preserved in a magnetic field and implies that the role of generalized Hubbard model interactions
at the neutrality point in Hartree-Fock theory is simply to shift all energy levels by an irrelevant constant.
This property is independent of the dependence of the interaction on site-separation.  
When translated to the continuum model, this property implies that 
for the case of a neutral graphene sheet both $\mathcal{E}$ in Eq.(~\ref{hmf}) should be independent of 
$n$ and equal to $\langle 0 | H_{e-e} | 0\rangle$ and that $\mathcal{B}_{x}$ should vanish.
Although the regularization procedure discussed above recovers this 
result with errors that vanish with 
cutoff like $1/\sqrt{M}$, the particle-hole symmetry property is so essential 
to the observed properties of graphene sheets that these errors are 
uncomfortably large at practical values of $M$.  We therefore propose the following 
regularization procedure for Hartree-Fock Landau level calculations for the massless 
Dirac model of graphene: i) Solve the Hartree-Fock equations for neutral graphene 
by setting $\Delta_{0,0}=1/2$, and $\Delta_{n,n}=(1-sgn(n))/2$ for $n \ne 0$.
ii) For the neutral case choose $\langle 0 | H_{e-e} | 0\rangle$ as the zero of energy 
and set $\mathcal{E}$ and $\mathcal{B}_{x}$ to zero to compensate for the
violation of particle-hole symmetry caused by a finite $M$ cut-off iii) In the case of 
charged graphene sheets $\Delta_{n',n}$ must be determined by a self consistent 
calculation in which $\mathcal{E}$ and $\mathcal{B}_{x}$ are replaced by
the the difference between their neutral and charged system values.
The Hartree-Fock energy levels of a neutral graphene sheet obtained 
by following i) are illustrated in Fig.(~\ref{figure1}).  The logarithmic 
dependence of all levels on cut-off $M$ in this figure is expected to 
appear experimentally as a weak logarithmic correction to the $\sqrt{B}$ 
dependence of Landau level energies on field expected for non-interacting electrons.  

This work was supported by the Welch Foundation and by the National 
Science Foundation under grant DMR-0606489.

\end{document}